# Device physics of van der Waals heterojunction solar cells

Marco M. Furchi*, Florian Höller*, Lukas Dobusch, Dmitry K. Polyushkin, Simone Schuler, and Thomas Mueller‡

*Vienna University of Technology, Institute of Photonics, Gußhausstraße 27-29, 1040 Vienna, Austria*

\* These authors contributed equally to this work.
‡ Corresponding author: thomas.mueller@tuwien.ac.at

**Heterostructures based on atomically thin semiconductors are considered a promising emerging technology for the realization of ultrathin and ultralight photovoltaic solar cells on flexible substrates. Much progress has been made in recent years on a technological level, but a clear picture of the physical processes that govern the photovoltaic response remains elusive. Here, we present a device model that is able to fully reproduce the current-voltage characteristics of type-II van der Waals heterojunctions under optical illumination, including some peculiar behaviors such as exceedingly high ideality factors or bias-dependent photocurrents. While we find the spatial charge transfer across the junction to be very efficient, we also find a considerable accumulation of photogenerated carriers in the active device region due to poor electrical transport properties, giving rise to significant carrier recombination losses. Our results are important to optimize future device architectures and increase power conversion efficiencies of atomically thin solar cells.**

INTRODUCTION

Two-dimensional (2D) semiconductors [1–5] provide a unique opportunity for the realization of ultrathin and ultralight photovoltaic solar cells [6], owing to their strong optical absorption in the solar spectrum region [3, 7], high internal radiative efficiencies [8], and favorable band gaps for both single-junction and tandem cells [9]. Theoretical estimates of power conversion efficiencies (PCEs) have predicted efficiency values exceeding 25 % [9], indicating that 2D semiconductors may become competitive with conventional photovoltaic technologies. The suitability of 2D materials for photovoltaic applications was first demonstrated in lateral p-n junctions [10–12], defined by split-gate electrodes, and in lateral Schottky junctions [13]. However, those device architectures do not allow for easy scalability of the photoactive area, for which a vertical junction would be desirable. Vertical van der Waals heterostructures [14] can be obtained by manual stacking [15, 16] or growth [17–19] of different 2D materials in a layered configuration. It has been shown that $MoS_2$ and $WSe_2$, when placed on top of each other, form a type-II heterojunction [20–24], with the lowest-energy conduction band states spatially located in the $MoS_2$ layer and the highest-energy valence band states in $WSe_2$. Relaxation of photogenerated carriers, driven by the conduction and valence band offsets, then results in a charge transfer across the 2D junction and a sizeable photovoltaic effect [20–22]. Similar results have been obtained using other material combinations that exhibit type-II band alignment, including $MoS_2/WS_2$ [25], $MoS_2$/black phosphorus [26, 27], $MoTe_2/MoS_2$ [28, 29], $GaTe/MoS_2$ [30], $MoSe_2/WSe_2$ [31, 32], $MoS_2$/carbon nanotubes [33], $MoS_2$/pentacene [34], $MoS_2$/silicon [35, 36], and many more. In addition, homojunction architectures have been explored, in which chemical doping is applied to form a vertical p-n junction in the same 2D material. Examples include plasma-induced p-doping of the upper layers in an n-type $MoS_2$ multi-layer crystal [37] and mechanical stacking of few-layer flakes of n-type $MoS_2$:Fe on top of p-type $MoS_2$:Nb [38]. In an optimized ∼15 nm thick $MoS_2/WSe_2$ heterostructure, an experimental absorbance of >90 %, an external quantum efficiency (EQE; the ratio between collected charge carriers and incident photons) exceeding 50 %, and a (single-wavelength) PCE of 3.4 % have been achieved [39].

The PCE, defined as the fraction of incident optical power $P_{opt}$ that is converted into electricity with output power $P_{el}$, is the most important parameter describing a photovoltaic device. It is given by the product of open-circuit voltage $V_{OC}$, short-circuit current $I_{SC}$, and fill factor $FF$:

$$PCE = \frac{P_{el}}{P_{opt}} = \frac{V_{OC} I_{SC} FF}{P_{opt}}. \qquad (1)$$

Today, fill factors in 2D heterostructure photovoltaic structures are typically in the range 0.3–0.5, only half as large as in conventional silicon solar cells. Closely connected to low fill factors are excessively high (≫2) ideality factors and low short-circuit currents, pointing towards substantial carrier recombination losses. Open-circuit voltages are typically less than 0.6 V, implying a band gap-$V_{OC}$ offset larger than 0.8 V [9]. For all these reasons, PCEs in 2D photovoltaic devices have as yet remained below 5 %, much lower then the Shockley-Queisser limit for their band gaps.

Besides technological challenges, the lack of a clear picture of the device physics in 2D heterostructure solar cells hampers further progress. Optimization of device architectures, however, will require an in-depth understanding of exciton dissociation, carrier transport processes, and recombination losses. Here, we address these questions by presenting a systematic experimental study of a $MoS_2$/$WSe_2$ van der Waals heterostructure and, based on the results, a model that reproduces the current-voltage characteristics under optical illumination. While we find the exciton dissociation to be very efficient, we also find a considerable pile-up of photocarriers in the device due to poor electrical transport properties, giving rise to carrier recombination and consequently low $FF$, $I_{SC}$ and $V_{OC}$-values. We finally provide guidelines to optimize future device layouts and increase PCEs.

**RESULTS**

**Van der Waals heterostructure solar cell.** Fig. 1a shows a schematic illustration of the $MoS_2$/$WSe_2$ heterostructure investigated in this work. Both layers exhibit monolayer thickness, as verified by Raman spectroscopy [40, 41]. An optical micrograph of the device can be found in Supplementary Fig. S1a and details about the fabrication process in the Methods section. The corresponding band diagram under short-circuit conditions is schematically depicted in Fig. 1b. As demonstrated later, the charge transfer from one layer into the other occurs with high efficiency. For the subsequent carrier transport to the contacts, driven by the lateral built-in field, it is hence justified to consider electron concentrations in $MoS_2$ (with quasi-Fermi level $E_{F,e}$) and hole concentrations in $WSe_2$ (with quasi-Fermi level $E_{F,h}$) only, as schematically depicted in Fig. 1d. For this reason, we refer to the $MoS_2$ sheet as electron transport layer (ETL) and the $WSe_2$ as hole transport layer (HTL). From the data reported in Ref. [42], we estimate an effective band gap of $E_{g,eff} = E_{ETL} - E_{HTL} = E_{CB,M} - E_{VB,W} \approx 1.3$ eV.

As reported previously [20], the electrical characteristics can be controlled by electrostatic doping via a back-gate voltage $V_G$, applied to the silicon substrate (Fig. 1c). For a large positive $V_G$ we find resistive n-n behavior, whereas for an appropriate choice of $V_G$ an atomically thin p-n junction is formed and the device current $I$ as a function of external bias voltage $V$ displays diode-like rectification behavior (inset in Fig. 1c, dashed line). All further measurements reported in this article were performed in the p-n regime, indicated by the arrows in Fig. 1c. Under optical illumination (solid line), the curve then passes through the fourth quadrant, meaning that electrical power can be extracted. The EQE is ~1 % under low-intensity (~1 kWm$^{-2}$) illumination at 532 nm wavelength and decreases for higher intensities. Compared to traditional solar cells, the device exhibits some unusual behaviors. For example, the photocurrent (PC), defined as $I_{\text{ph}} = -(I_{\text{illum}} - I_{\text{dark}})$, changes sign and becomes negative at $V \gtrsim V_{\text{OC}}$. $I_{\text{ph}}$ also does not become fully saturated under reverse bias, which is an indication for insufficient carrier extraction. These characteristics have been consistently observed not only in all devices that we investigated, but also in many other van der Waals heterostructures reported in literature.

**Transport model.** The current-voltage $I(V)$ characteristic of a photovoltaic solar cell is usually described by the Shockley diode equation [43]

$$I(V) = I_0 \left[\exp\left(\frac{qV}{n_{\text{id}}k_B T}\right) - 1\right] - I_G, \qquad (2)$$

where $I_0$ is the dark generation current, $q$ the elementary charge, $k_B$ Boltzmann's constant, and $T$ the temperature. $I_G$ denotes the photogenerated current and $n_{\text{id}}$ is the ideality factor whose value depends on the type of recombination mechanism: $n_{\text{id}} = 1$ for direct bimolecular (Langevin) recombination and $n_{\text{id}} = 2$ for trap-mediated (Shockley-Read-Hall) recombination. Under open-circuit conditions ($V = V_{\text{OC}}$) the current becomes zero and the well-known relation $\partial V_{\text{OC}}/\partial \ln(P_{\text{opt}}) = n_{\text{id}}k_B T/q$ is derived, from which the ideality factor can be determined. Fig. 2a shows $I(V)$ characteristics of our device at different optical excitation powers (symbols) from which we can readily extract the open-circuit voltages $V_{\text{OC}}$. Together with the incident optical power $P_{\text{opt}}$, we then determine an ideality factor of $n_{\text{id}} = 1.6$ (see Supplementary Fig. S2), which indicates an involvement of both recombination mechanisms [20, 21]. However, if we plot equation (1) with $n_{\text{id}} = 1.6$ in Fig. 2a (dash-dotted line; $P_{\text{opt}} = 16$ nW) we find very poor agreement with the experimental data. The Shockley equation strongly overestimates short-circuit current, fill factor, and forward current. This is in contrast to lateral 2D semiconductor p-n junctions defined by split-gate electrodes, that can be well described by the Shockley model [10, 11].

To obtain better modeling of solar cells a series (contact) resistance is often taken into account. However, as shown in see Supplementary Fig. S3, an extended Shockley model does not either fit our data. Particularly, the strong illumination dependence of the forward current and the interception of the $I(V)$ curves cannot be explained. Another mechanism that can affect the electrical characteristics of solar cells is the build-up of space charge regions at the contacts or in the bulk, as a result of strongly unbalanced electron and hole transport [44]. In order to explore this mechanism, we plot in Fig. 2b the current versus compensation voltage $V_{\text{OC}} - V$ on a double-logarithmic scale (symbols). The current shows the usual linear dependence at low compensation voltages, due to the competition between drift and diffusion of photogenerated carriers, and then a smooth transition to the saturation regime for larger voltages. The characteristic fingerprint of space charge-limited transport [44] – a region with square-root dependence $I \propto (V_{\text{OC}} - V)^{1/2}$ – is absent. Hence, we conclude that the carrier transport in our device is balanced. A third mechanism that can affect the $I(V)$ characteristics of diodes are interface inhomogeneities. A prominent example are barrier height variations in Schottky diodes [45], that can result in $n_{\text{id}} > 2$. We rule out this possibility for the following reasons. First, the forward current in atomically thin p-n junctions is of different origin than in conventional diodes; it is governed by tunneling-mediated interlayer recombination, rather than carrier injection over a potential barrier [20, 21]. Second, sample inhomogeneities cannot explain the illumination dependent device behavior.

In the following we will instead argue that the photovoltaic response is transport-limited. It is thus inappropriate to employ Shockley's model, as it does not account for the impact of charge transport (it assumes infinitely large conductivities for electrons and holes). In order to obtain better modelling of the $I(V)$ characteristics we follow the approach by Würfel *et al.*, initially developed for organic solar cells [46]. In brief, carrier accumulation due to poor transport properties leads to a quasi-Fermi level splitting $qV_{\text{int}} = E_{\text{F,e}} - E_{\text{F,h}}$ in the electron and hole transport layers that differs from the externally applied voltage:

$V_{\text{int}} = V - \xi I/\sigma$, where $\sigma$ is the electrical conductivity under optical illumination and $\xi$ is a geometry factor with unit $\text{m}^{-1}$. It is only for open-circuit conditions that $V_{\text{int}} = V$, and the Shockley equation (2) can be employed. For all other biases, where $I \neq 0$, it has been suggested [46] to use $V_{\text{int}}$ in (2) instead of $V$. A closed form approximation of the $I(V)$ curves can then be derived [47], that has been shown to reproduce the results of full drift-diffusion simulations for a wide range of parameters:

$$I(V) = I_G \left\{ \exp\left[\frac{q(V - V_{\text{OC}})}{k_B T(1 + \alpha)}\right] - 1 \right\}, \quad (3)$$

with a dimensionless figure of merit $\alpha$, that is a direct measure of non-ideal device behavior as a result of insufficient carrier extraction. It is given by

$$\alpha = \xi \frac{q}{k_B T} \frac{I_G}{\sigma_i} \exp\left(-\frac{qV_{\text{OC}}}{2k_B T}\right), \quad (4)$$

where $\sigma_i$ denotes the intrinsic electrical conductivity in the dark (see Supplementary Note S1 and Ref. 47). If we fit equation (3) to the experimental data (solid lines in Figs. 2a and b) we find excellent agreement for all illumination intensities. The photogenerated current $I_G$ scales linearly with $P_{\text{opt}}$, as expected (see Supplementary Fig. S4). The ideality parameter $\alpha$ varies over a wide range and reaches values as high as ~78. Insufficient carrier extraction, described by large $\alpha$ values, leads to high carrier densities and consequently to a large quasi-Fermi level splitting $V_{\text{int}}$, even under short-circuit conditions. As the interlayer recombination is governed by $V_{\text{int}}$, this results in recombination losses and $I_{\text{SC}} < I_G$. Under forward bias, the accumulated charge enhances the conductivity, resulting in the experimentally observed crossing of dark and illuminated $I(V)$ curves. As shown in Supplementary Note S2, expression (4) can be further simplified to yield

$$\alpha = K \frac{\sqrt{P_{\text{opt}} k_{\text{rec}}}}{T\mu}, \quad (5)$$

where $\mu$ denotes the effective carrier mobility in the electron and hole transport layers, $k_{\text{rec}}$ is the interlayer recombination coefficient, and all other physical constants and geometry factors have been lumped into the prefactor $K$. The expression predicts a square-root dependence of $\alpha$ on the optical power, which is indeed observed experimentally (Fig. 3a).

The question that remains to be addressed is why the carrier extraction in our device is inefficient, given the rather high mobility of 2D semiconductors (typically 10–100 cm$^2$/Vs). In high-mobility materials, such as crystalline silicon, $\alpha$ approaches zero and the impact of charge transport is negligible. In organic materials, on the other hand, $\mu$ is extremely low and transport-limitations are expected. From the dark $I(V)$s in Fig. 1c it is apparent that varying the sample temperature provides us with an opportunity to tune $\sigma_i$ ($\propto I_{\text{dark}}$) in the p-n regime over almost three orders of magnitude. In Fig. 3b (blue symbols) we depict its temperature variation, and we find that it can be described by a thermally activated transport model [48], $\sigma_i(T) \propto \exp(-E_a/k_B T)$, with an activation energy $E_a \approx 80$ meV (dashed line). Next, we record the temperature dependent photovoltaic properties (Fig. 2c). From that we determine $\alpha(T)$ and then, with equation (4), the temperature dependence of $\sigma_i$. If we plot the results as red symbols in Fig. 3b, we find a striking similarity with the dark current measurements. The limitation of the present device architecture becomes clear now. For efficient carrier injection/extraction the heterojunction has to be operated in the p-n regime, in which both layers are strongly depleted (sub-threshold regime). Transport of photogenerated carriers then occurs via thermal activation from the impurity band tails (intrinsic or induced by disorder) [49–51], which results in low effective carrier mobility and, hence, recombination losses.

As the parameter $\alpha$ determines the shape of the $I(V)$ curves, it is a good figure of merit to assess the quality of solar cells. Neher *et al.* suggested an empirical expression that relates $\alpha$ to the fill factor [47]. In Fig. 3c we plot this expression for the range of open-circuit voltages that is relevant in this work. In the same plot we summarize our experimental results as circular symbols, where the fill factor was determined, as usual, from the maximum power point: $FF = \max(P_{\text{el}})/(V_{\text{OC}} I_{\text{SC}})$. We conclude that higher conductivities and lower illumination intensities result in better device performance.

**Interlayer charge transfer.** So far, we have assumed an ultrafast interlayer charge transfer (exciton dissociation), and subsequent charge transport in the ETL and HTL on a longer time scale. We now substantiate this claim by presenting PC autocorrelation measurements, a powerful technique to study the carrier dynamics in 2D materials [52, 53] and heterostructures [54]. It exploits the non-linearity of the photoresponse to infer the underlying dynamics of the system. Using this technique, the device is optically excited with a pair of femtosecond laser pulses (0.2 ps pulse duration), and the integrated PC is recorded as a function of the temporal delay between the pulses. Experimental details are presented in the Methods section and Supplementary Figs. S6 and S7.

Fig. 4a shows the absorbance of our device, as determined by a Kramers-Kronig analysis [55] of the reflectance spectrum. We identify absorption peaks at 1.65, 1.89 and 2.03 eV, that can be associated with the excitonic ground-state transition of $WSe_2$ and the A- and B-excitons in $MoS_2$, respectively. We set the excitation energy of our pulsed laser to 1.65 eV (laser spectrum in Fig. 4a) to resonantly excite excitons in the $WSe_2$ layer only. The resulting PC thus originates from the electron transfer from $WSe_2$ into $MoS_2$. Fig. 4b depicts the power dependence, where above ~0.5 µW average incident power a clear saturation behavior is observed. The non-linearity might either stem from phase-space filling, that leads to absorption saturation, or from many-body interactions, such as exciton-exciton annihilation or Auger recombination. Fig. 4c shows autocorrelation traces taken for different pulse energies (symbols). Below saturation (bottom curve), a symmetric trace is obtained with (undersampled) optical interference fringes around zero time delay. The linear photoresponse corresponds to the interferometric first-order autocorrelation of the incident light and does not provide any insight in dynamical device behavior. However, when the pulse energy is increased to exceed the saturation threshold, the autocorrelation data become increasingly asymmetric. The characteristic time constant of the nonlinear background signal can be associated with the exciton dissociation time (see Supplementary Note S3). This time is shorter than the pulse duration and we conclude that the dissociation occurs within less than 0.2 ps, similar to what has been obtained in all-optical pump/probe experiments [56]. The charge transfer on a time scale shorter than characteristic transport times, confirms the validity of the transport model in Fig. 1d. It also suggests an efficient charge separation from bound excitons after absorption, despite the large exciton binding energy in 2D materials.

**DISCUSSION**

2D semiconductors contain a significant amount of electronic band tail states [49–51]. As shown in the previous section, these defects trap charge and adversely affect the electronic transport and, consequently, the PCE. We conclude that the performance of van der Waals heterojunction solar cells is transport-limited and we suggest that equation (3) should be used to describe the photovoltaic response, instead of the commonly used Shockley equation. An investigation of the role of defects in the limitation of the open-circuit voltage is beyond the scope of this article, but we note that it is well understood that defects not only reduce $FF$ and $I_{SC}$, but also $V_{OC}$ [57].

To further test out model, we analyzed some exemplary cases from literature. The symbols in Fig. 5a show data points extracted from Ref. 21, along with a fit of equation (3) as solid line. This work employed a device structure similar to the one presented in this article, but with more favorable (unintentional) doping of the $MoS_2$ and $WSe_2$ layers. This resulted in a comparably high $FF$ despite the high illumination intensity used in the experiment (diamond-shaped symbol in Fig. 3c). We believe that transport-limitations also occur in device architectures that employ optically transparent (graphene or indium-tin-oxide) electrodes for vertical carrier extraction. There, the charge transport occurs on much shorter length scales, but also with extremely low out-of-plane mobilities (~$10^{-2}$ cm$^2$/Vs) [54]. In Figs. 5b and c we present two examples taken from Refs. 39 and 37, respectively. The data can again be well fitted by equation (3) and the results are summarized as yellow symbols in Fig. 3c. Interestingly, lateral p-n junctions, based on split-gate electrodes, typically show better photovoltaic properties ($V_{OC}$ > 0.85 V [10], $FF$ > 0.7 [58], ideality factor < 2 [11]) than van der Waals structures. This is explained by the higher electrical conductivities in these devices, because of independent doping of the p- and n-type regions.

Based on our results, we finally provide guidelines that might allow to avoid charge pile-up in future device architectures and harness the true potential of 2D materials in photovoltaic applications. We suggest (i) to increase the device conductivity, e.g. by elimination of charge traps or by (chemical) doping, and (ii) to make the active region as thin/short as possible to facilitate more efficient charge extraction before recombination.

## METHODS

**Device fabrication.** The van der Waals heterojunction device was fabricated by stacking of mechanically exfoliated $MoS_2$ and $WSe_2$ flakes on a $SiO_2$/silicon substrate, as described in our previous work [20]. 2D semiconductor monolayer thicknesses were verified beforehand by Raman spectroscopy and photoluminescence (see Supplementary Figs. S1b and S5). Palladium/gold contact electrodes were defined using electron-beam lithography and metal evaporation. Palladium was chosen as trade-off between forming an n-type contact to $MoS_2$ and a p-type contact to $WSe_2$. After fabrication, the samples were annealed in vacuum (~5×10$^{-6}$ mbar) for several hours at a temperature of $T = 380$ K and mounted on a chip holder.

**Optical measurements.** The photovoltaic response was measured using a 532 nm wavelength laser, that was focused to a ~1 μm full-width-at-half-maximum (FWHM) spot on the device. The sample was mounted on a motorized XYZ-stage for precise alignment with the laser spot. A camera allowed us to view the illumination position on the device. The optical excitation power was adjusted using a variable optical attenuator and the electrical characteristics were acquired using a Keithley source-meter. Room-temperature measurements were performed under ambient conditions and low-temperature measurements in a flow-cryostat.

**Time-resolved measurements.** In time-resolved photocurrent measurements, a sequence of two ultrashort laser pulses ($\tau_p = 0.2$ ps FWHM pulse duration, $f_R = 76$ MHz repetition rate) was generated using a wavelength-tunable Ti:Sapphire laser and a Michelson interferometer. The time delay between the two pulses was adjustable via a computer-controlled mechanical translation stage in one of the interferometer arms. The laser pulses were then focused with a microscope objective to a ~0.7 μm diameter (FWHM) spot on the sample. Time-resolved experiments were performed at $T = 300$ K and under ambient conditions.

**Data availability:** The data used in this study are available upon request from the corresponding author.

**Acknowledgments:** We thank Karl Unterrainer for providing access to a Ti:Sapphire laser and acknowledge financial support by the Austrian Science Fund FWF (START Y 539-N16) and the European Union (grant agreement No. 696656 Graphene Flagship).**Author contributions:** T.M. conceived the experiment. M.M.F. and L.D. fabricated the samples. F.H. and M.M.F. carried out the electrical and optical measurements. D.K.P. and S.S. assisted with the measurements and the sample fabrication. T.M., M.M.F. and F.H. analyzed the data. T.M. prepared the manuscript. All authors discussed the results and commented on the manuscript.

## ADDITIONAL INFORMATION

**Supplementary information** accompanies the paper on the npj 2D Materials and Applications website.

**Competing financial interests:** The authors declare no competing financial interests.

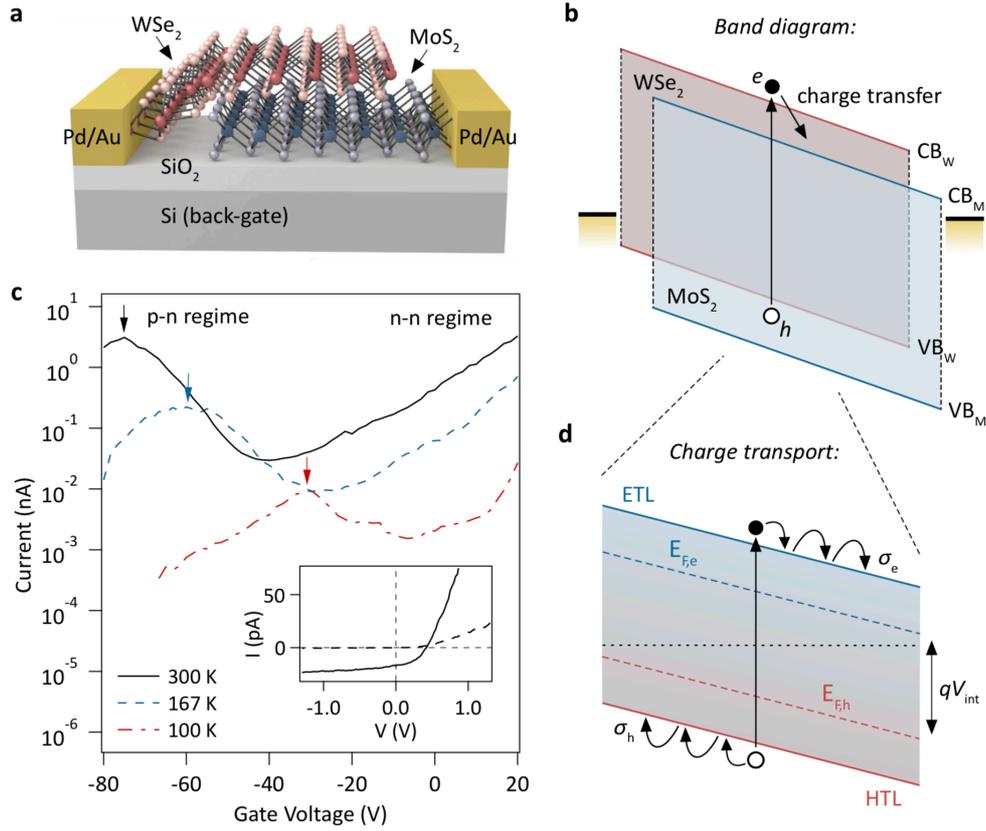

**Fig. 1 | Van der Waals heterostructure solar cell. a,** Schematic illustration of the device structure. **b,** Band diagram in lateral direction under short-circuit conditions. CB, conduction band; VB, valence band. **c,** Temperature dependent device current, recorded by scanning the back-gate voltage $V_G$. The arrows indicate the p-n regime, in which the device shows a photovoltaic response. Inset: Current-voltage characteristic at room-temperature, recorded at $V_G$ = -75 V. Dashed line: dark current; solid line: under optical illumination. **d,** Charge transport model, with effective band gap $E_{g,\text{eff}} = E_{\text{CB,M}} - E_{\text{VB,W}}$. The electron concentration in MoS$_2$ is described by the quasi-Fermi level $E_{F,e}$ and hole concentration in WSe$_2$ by $E_{F,h}$. The quasi-Fermi level splitting, expressed as voltage, is $qV_{\text{int}}$. $E_{F,e}$ and $E_{F,h}$ are assumed to exhibit the same constant tilt throughout the entire device. ETL, electron transport layer; HTL, hole transport layer. Electron and hole conductivities in the ETL and HTL are denoted as $\sigma_e$ and $\sigma_h$, respectively. Note, that in an ideal solar cell with infinite carrier mobility, $V_{\text{int}}$ is zero under short-circuit conditions (illustrated by the black dotted line).

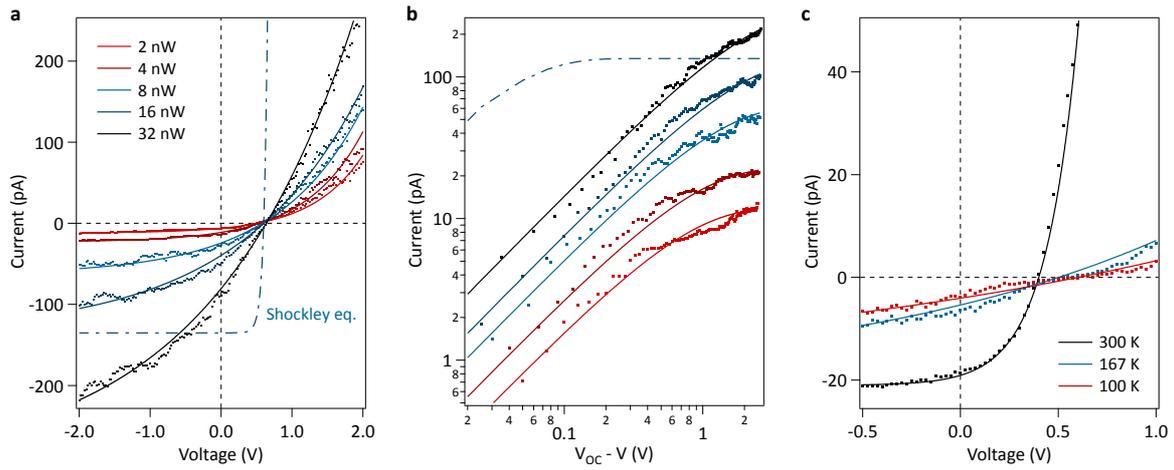

**Fig. 2 | Current-voltage characteristics. a,** Illumination power dependent $I(V)$ curves at room temperature. Symbols: experimental data; solid lines: fit of equation (2); dash-dotted line: fit of Shockley equation to the data recorded with $P_{\text{opt}} = 16$ nW. **b,** Same data as in a, but plotted versus compensation voltage $V_{\text{OC}} - V$ on a double-logarithmic scale **c,** Temperature dependent $I(V)$ characteristics.

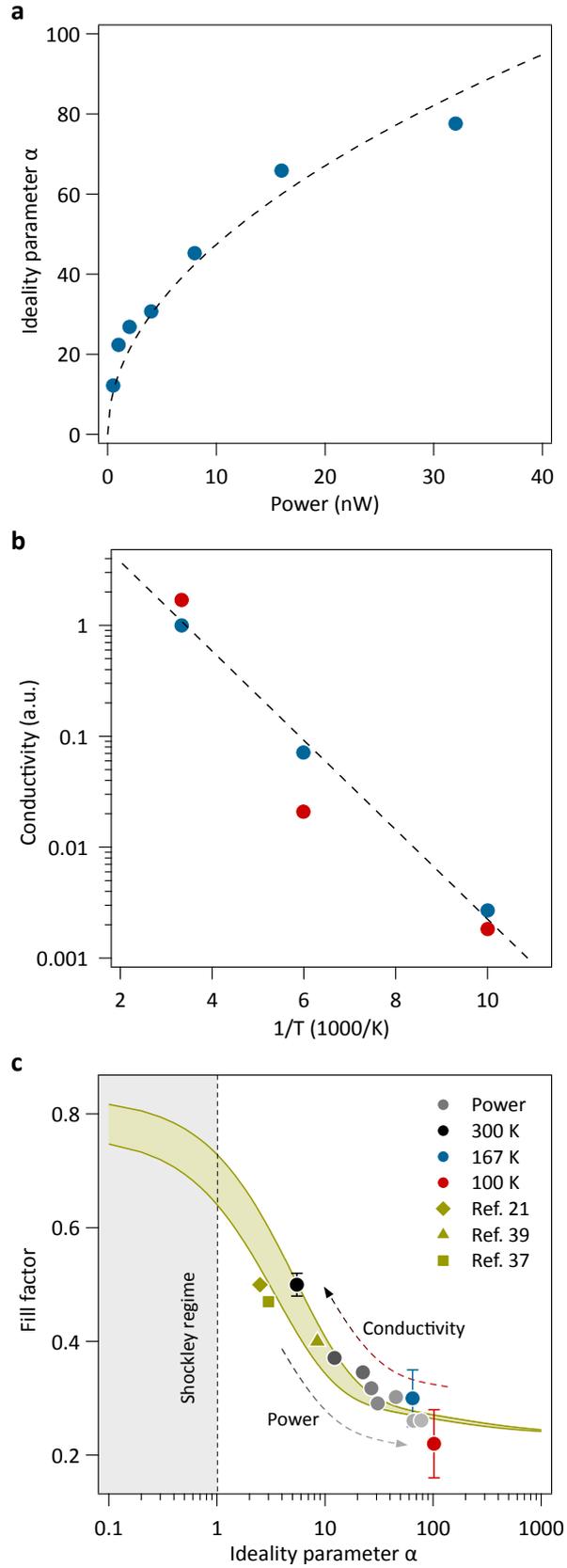

**Fig. 3 | Device non-ideality due to transport-limitations. a,** Parameter $\alpha$ extracted from the $I(V)$ curves versus incident optical power (symbols). Dashed line: fit of equation (5). **b,** Arrhenius plot of (normalized) conductivity $\sigma_i$, as extracted from dark-current

measurements (red symbols) and photovoltaic measurements (blue symbols). Dashed line: thermally activated transport model. **c,** Circular symbols: *FF* versus *α* for all measurements presented in this article. Yellow symbols: application of our model to the data in Refs. 21, 37 and 39. Yellow area: plot of the empirical equation from Ref. 47 for open-circuit voltages between 0.4 and 0.65 V.

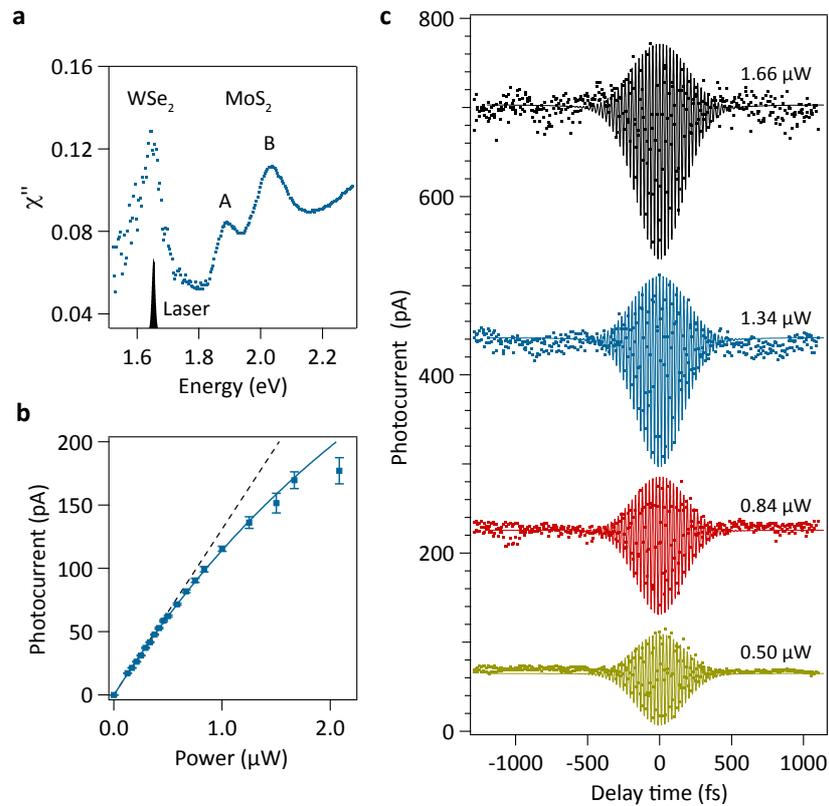

**Fig. 4 | Time-resolved photocurrent measurement. a,** Absorbance spectrum of the heterostructure (symbols) and femtosecond laser spectrum (black line). **b,** Power dependence of the PC (symbols: measurement, solid line: theoretical model). Above ~0.5 µW a clear deviation from an unsaturated behavior (dashed line) is observed. **c,** PC autocorrelation traces for different pulse energies (symbols: measurement, lines: theoretical model). The curves are offset for clarity.

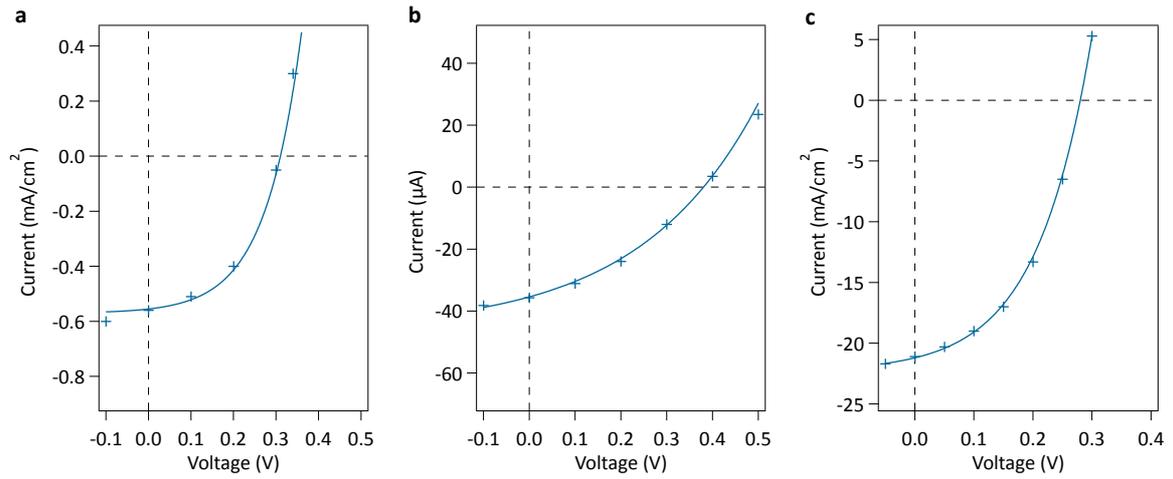

**Fig. 5 | Fitting of data from other publications.** Symbols: data points extracted from publications; lines: fit of equation (3). **a,** Lee *et al.* [21]. **b,** Wong *et al.* [39]. **c,** Wi *et al.* [37].